\input epsf
\documentstyle[12pt]{article}

\newlength{\dinwidth}
\newlength{\dinmargin}
\newcommand{\resection}[1]{\setcounter{equation}{0}\section{#1}}

\begin{document}

\begin{center}
  \begin{Large}
  \begin{bf}
UNIVERSIT\'E DE GEN\`EVE \\
  \end{bf}
  \end{Large}
\smallskip
  \begin{small}
SCHOLA GENEVENSIS MDLIX
  \end{small}
\end{center}
\goodbreak
\begin{center}
\bigskip\vbox
{
\vskip 3truecm
\noindent
\includegraphics{unige.ps}
\vskip 2truecm
\noindent
}
\end{center}
\bigbreak
\begin{center}
 \begin{Large}
  \begin{bf}
SOME RESULTS ON THE BESS MODEL AT FUTURE COLLIDERS$^*$\\
 \end{bf}
  \end{Large}
  \vspace{5mm}
  \begin{large}
R. Casalbuoni$^{(a,b)}$, P. Chiappetta$^{(c)}$, A. Deandrea$^{(d)}$,
 S. De Curtis$^{(b)}$,\\
D. Dominici$^{(a,b)}$ and R. Gatto$^{(d)}$\\
  \end{large}
  \vspace{5mm}
$(a)$ Dipartimento di  Fisica, Univ. di Firenze, I-50125 Firenze, Italy\\
$(b)$ I.N.F.N., Sezione di Firenze, I-50125 Firenze, Italy\\
$(c)$ CPT, CNRS, Luminy Case 907, F-13288 Marseille, France\\
$(d)$ D\'ept. de Phys. Th\'eor., Univ. de Gen\`eve, CH-1211 Gen\`eve 4\\
  \vspace{5mm}
\end{center}
\vspace{1.5cm}
\begin{center}
UGVA-DPT 1995/02-883\\
hep-ph/9502325
\end{center}
\vspace{3cm}
\noindent
$^*$ Partially supported by the Swiss National Foundation. Contribution to
the ESB \& BSM Working group of DPF Long Range Planning Study, Albuquerque,
August 7, 1994.
\newpage
\def\lq{\left [}
\def\rq{\right ]}
\def\qq{Q^2}
\def\dmu{\partial_{\mu}}
\def\dmus{\partial^{\mu}}
\def\AA{{\cal A}}
\def\BB{{\cal B}}
\def\Tr{{\rm Tr}}
\def\gp{g'}
\def\gs{g''}
\def\rs{\sqrt{s}}
\def\ggs{\frac{g}{\gs}}
\def\mpp{M_{P^+}}
\def\mpm{M_{P^-}}
\def\mpt{M_{P^3}}
\def\mpz{M_{P^0}}
\newcommand{\be}{\begin{equation}}
\newcommand{\ee}{\end{equation}}
\newcommand{\bea}{\begin{eqnarray}}
\newcommand{\eea}{\end{eqnarray}}
\newcommand{\nn}{\nonumber}
\newcommand{\dd}{\displaystyle}
\setcounter{page}{1}
\font\twelve=cmbx10 at 15pt
\font\ten=cmbx10 at 12pt
\font\eight=cmr8
\def\build#1_#2^#3{\mathrel{\mathop{\kern 0pt#1}\limits_{#2}^{#3}}}
\def\vect#1{\overrightarrow{#1\kern 1pt}\kern-1pt}
\newcommand{\rf}[1]{(\ref{#1})}
\newcommand{\beq}[1]{\begin{equation}\label{#1}}
\newcommand{\eeq}{\end{equation}}
\newcommand{\beqa}[1]{\begin{eqnarray}\label{#1}}
\newcommand{\eeqa}{\end{eqnarray}}
\newcommand{\beqan}{\begin{eqnarray*}}
\newcommand{\eeqan}{\end{eqnarray*}}
\def\wt{\widetilde}
\def\wh{\widehat}
\def\ul{\underline}
\def\ol{\overline}
\def\tg{\mathop{\rm tg}\nolimits}
\def\cotg{\mathop{\rm cotg}\nolimits}
\arraycolsep2pt

\resection{Introduction}

We present some results on the usefulness of upgraded Tevatron, LHC
proton proton collider and linear $e^+e^-$ colliders in the TeV range
to test the idea of a strongly interacting sector as responsible for
the electroweak symmetry breaking.

The calculations are performed within an effective
lagrangian description, called the BESS model, which provides for a
rather general frame based on the
standing point of custodial symmetry and gauge invariance, without
specifying any dynamical scheme.

We are interested in studying a spontaneous symmetry breaking
avoiding physical scalar particles, i.e. in a non linear rather than
in a linear way. An effective lagrangian describing in an unified way
mass terms and interactions of the standard electroweak gauge bosons
has been derived~  \cite{BESS} as a gauged non linear
$\sigma$-model.

Using extensively the fact that any non linear
$\sigma$-model is gauge equivalent to theories with additional
hidden local symmetry~\cite{bando}, we can build up new vector
resonances, similar to ordinary $\rho$ vector mesons or to
the techni-$\rho$ particle of technicolor theories, as the
gauge bosons associated to the hidden symmetry group of $SU(2)$
type. Under the assumption they are dynamical, we will get the
$SU(2)$ BESS lagrangian~\cite{BESS} (BESS standing for Breaking
Electroweak Symmetry Strongly).

Therefore the minimal BESS model
is described by a Yang Mills lagrangian whose gauge group is $SU(2)_L
\otimes U(1)_Y \otimes SU(2)_V$

\beqa{1}
& {\cal L}_{BESS} = & - \frac{v^2}{4}
\left[ Tr (W - B)^2 + \alpha \ Tr (W + B - V)^2 \right]
\nonumber \\[1mm]
& &
+ \frac{1}{2 g^2} Tr
\left( F_{\mu \nu} (W) F^{\mu \nu} (W) \right)
+ \frac{1}{2 g'^2} Tr
\left( B_{\mu \nu} B^{\mu \nu} \right)
\nonumber \\[1mm]
& &
+ \frac{2}{g''^2} Tr
\left( F_{\mu \nu} (V) F^{\mu \nu} (V) \right)
\eeqa
with $F_{\mu \nu} (V) = \partial_{\mu} V_{\nu} -
\partial_{\nu} V_{\mu} +
[V_{\mu},\ V_{\nu}]$ where $V_{\mu} = i \frac{g''}{2}
\frac{\tau^a}{2} V^a_{\mu},~(a=1,2,3)$, $\alpha$ is an arbitrary parameter
and $g''$ the new gauge coupling constant.
 The first term
within brackets is the usual mass term appearing in the Standard Model
(hereafter denoted as SM).

The theory is invariant under $ U (1)$ electromagnetic and $S U(2)$
custodial symmetry. The additional parameters it contains are
the mass $M_V$ (which depends on $\alpha$)
of the new bosons forming a degenerate triplet and
their gauge coupling constant $g''$.

The new particles are naturally coupled to fermions through mixing
between $\smash{\vect W}$ and $\smash{\vect V}$, although a direct
coupling, specified by a new parameter $b$, is possible.
The SM is recovered in the limit $g'' \to \infty$ and
$b = 0$. Mixings of the ordinary gauge bosons to the $V$'s are of
 the order of ${\cal O}
\left( {g}/{g''} \right)$. Due to these mixings, $V$ bosons are
coupled to fermions even for $b = 0$. Furthermore these couplings are
still present in the limit $M_V \to \infty$, and therefore the new
gauge bosons effects do not decouple in the large mass limit.

In the vector dominance approximation the BESS model corresponds to  a
technicolor model~\cite{farhi} with a single technidoublet.
If a non zero direct coupling of $V$ to fermions exists it corresponds
to an extended technicolor. The model can also incorporate axial vector
resonances by enlarging the additional gauge group from
$SU(2)_V$ to $SU(2)_L \otimes SU(2)_R$ local.

In what follows we will restrict ourselves to new vector bosons. Their
existence will indirectly manifest at LEP through deviations from SM
expectations~\cite{self}. For this purpose a low energy effective theory
valid for heavy resonances recently derived~\cite{anich} is useful.

\resection{Present limitations}

The analysis of LEP data, concerning the total width, the hadronic
and the leptonic width, the leptonic and bottom forward-backward
asymmetries, the $\tau$-polarization together with the cesium atomic parity
violation and the ratio $M_W / M_Z$, uses available full one loop
radiative correction programs.
We assume for the BESS model the same one-loop radiative corrections as for the
SM in which the Higgs mass is used as a cut-off $\Lambda$,
this brings in a dependence
on $\alpha_s,\ m_{top}$ and $\Lambda$.
We will rewrite the observable
quantities in terms  of the parameters
$\varepsilon_i$~\cite{alta}.

The BESS contribution reads:
\beqa{2}
\varepsilon_1
& = & \varepsilon_2 = 0 \nonumber \\[1mm]
\varepsilon_3
& = & \left( \frac{g}{g''} \right)^2 - \frac{b}{2}
\eeqa

This shows explicitly that through LEP data we are only sensitive
to one combination of BESS parameters i.e. $\varepsilon_3$. The
allowed region at $90\%$ CL in the $\left(
b, {g}/{g''}
\right)$ plane is shown in Fig.~1 for a top mass value of $174\pm 17$ GeV
and in the limit $M_V>>M_W$.
The chosen experimental value~\cite{alta2}
\beq{3}
\varepsilon_3^{\exp} = (3.9 \pm 1.7) 10^{-3}\eeq
corresponds to the latest LEP1 data combined with UA2/CDF/D0 ones
presented at the Glasgow conference, and we have added to (2.1) the
contribution coming from the radiative corrections \cite{alta2}
for $M_H=\Lambda=1~TeV$ and $m_{top}=174\pm 17$ GeV, which is \break
$\varepsilon_3^{\rm {rad.~corr.}}=(6.39^{-0.14}_{+0.20}) 10^{-3}$.

We further observe that, within the BESS model, we can explain
the two standard deviations from SM expectation for the $Z$
partial width in $b \bar b$
by assuming a sizeable
non zero direct coupling $b^\prime$ only for the heaviest generation
(as expected from one loop BESS radiative corrections proportional
to $m_f$).
After adding the SM expectation for $m_{top} = 170$ GeV to the BESS
model contribution $\varepsilon_b=-b^\prime/2$,  we get at 90\%
CL
\beq{4}
- 2.7\ 10^{- 2} \le b^\prime \le  0.32\ 10^{- 3}.
\eeq
 from $\varepsilon_b^{\exp} = (0.2 \pm 4.0) 10^{-3}$.

\smallskip
\centerline{
\epsfxsize=8truecm
\epsffile[69 263 498 700]{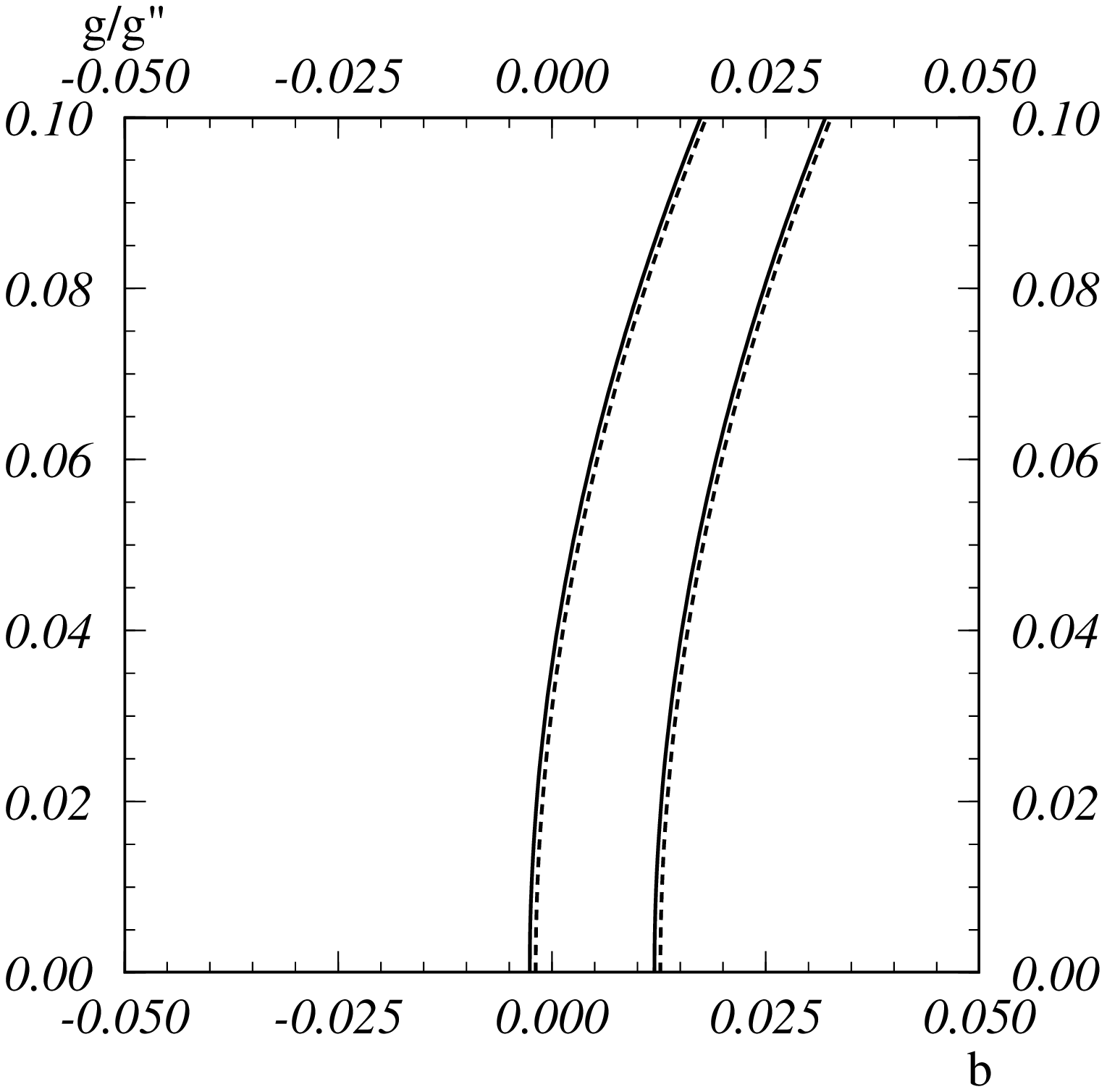}}

\baselineskip=10pt
\smallskip
\noindent
{\bf Fig. 1} - {\it  $90\%$ C.L. contour in the plane $(b,g/\gs)$
 from the measurement of $\varepsilon_3$. The solid (dashed) line is for
$m_{top}(GeV)= 191(157)$, $\Lambda=1~TeV$ and $\alpha_s=0.118$.}
\smallskip

\baselineskip=14pt

\resection{BESS at LHC}

At hadron colliders, as far as detection of a signal from a strongly
interacting symmetry breaking sector is concerned, vector boson pair
production is particularly relevant.
In the BESS model there are two main processes which compete for the production
of a pair of ordinary gauge bosons at a $pp$ collider: $q\bar q$ annihilation
and ordinary gauge boson fusion.
In the first mechanism a quark-antiquark pair annihilates into a $V$
 vector boson,
which then decays into a pair of ordinary gauge vector bosons. We stress
the fact that this process  is always operating in BESS independently of
the existence of a direct coupling of $V$ to fermions, because of the mixing.

We further observe that for masses of $V$ in the $TeV$ range, the $V$ decay
is dominated by the $WW$ and $WZ$ channels due to the large coupling of
the $V$ to the longitudinal components of the standard gauge bosons.
For this reason, in the computation of the width $\Gamma_V$
we have ignored the contribution from the fermionic channels since
it turns out to be completely negligible.

The second mechanism to produce $W/Z$ pairs is the rescattering of a pair
of ordinary gauge bosons, each being initially emitted from a quark or
antiquark leg. In BESS the rescattering process is naturally strong. In fact
the scattering of two longitudinally polarized $W/Z$'s proceeds via the
exchange of  a $V$ vector boson with large couplings at each vertex.
This process has been evaluated by using the effective-$W$ approximation
for the initial $W/Z$ and using the equivalence theorem for the
rescattering amplitudes.

It turns out that
the $pp\to W^\pm Z+X$ reaction is the most interesting
one in the framework of the BESS model. The
process $pp\to W^+ W^- +X$ is expected to suffer from a very severe
background coming from $pp\to t\bar t+X$, with $t$ and $\bar t$
both decaying into $W$.
Final leptonic configurations from $t\bar t$ production might also
simulate configurations from $W^\pm Z$, but the $Z$ mass reconstruction
and lepton isolation requirements
should protect from such a background \cite{josa}.
 The $ZZ$ mode has not been considered
because it does not proceed via an $s$-channel contribution in BESS.

The relevant backgrounds are the standard
model production of $W^\pm Z$ through quark-antiquark annihilation,
$\gamma W^\pm$ fusion and $W_T^\pm Z_T$ fusion.

 In our calculation we have made use of the
DFLM structure functions, for $\Lambda_{QCD}=260~MeV$.
For  the case of
the fusion process as for  $q\bar q$ annihilation we have
taken an evolution scale for the structure
functions equal to the
square of the invariant mass of the produced gauge boson pair.

We  assume LHC running at 14 $TeV$ with
a luminosity of  $10^{34}~cm^{-2}~sec^{-1}$.

Concerning the cuts, a first one
on the rapidity $y_{W,Z}$ of the final $W$ and $Z$, $|y_{W,Z}|\le 2.5$
was imposed to all cases.
Then we applied a lower
cut in $M_{WZ}$ (the
invariant mass of the $WZ$-pair),
approximately corresponding to the beginning
of the resonance at the left of the peak. An upper
cut has been fixed once for all at $M_{WZ}=3~TeV$, where
the resonance tail is already extinguished.
Finally a cut in $p_T$ (the transverse momentum of the $Z$) has been
obtained from the requirement of maximizing the statistical significance of
the signal, $S/(S+B)^{1/2}$, $S$ being the signal and $B$ the background.

The calculated event rates are largely observable
at the projected LHC energy and luminosity, for reasonable ranges of the
BESS parameters. Detailed studies of background and statistical significance
of signals versus background can be found in \cite{LHC} where the
$pp$ center of mass energy was assumed to be 16 $TeV$. The rate of the events
decreases by roughly 20$\%$ if we consider the presently planned energy
of 14 $TeV$ \cite{pseudo}.

As it is clear from the tables in ref. \cite{LHC},
the $q\bar q$ mechanism is found to be in general
dominant at LHC with respect to the fusion mechanism.

\smallskip
\centerline{
\epsfxsize=8truecm
\epsffile[45 214 551 654]{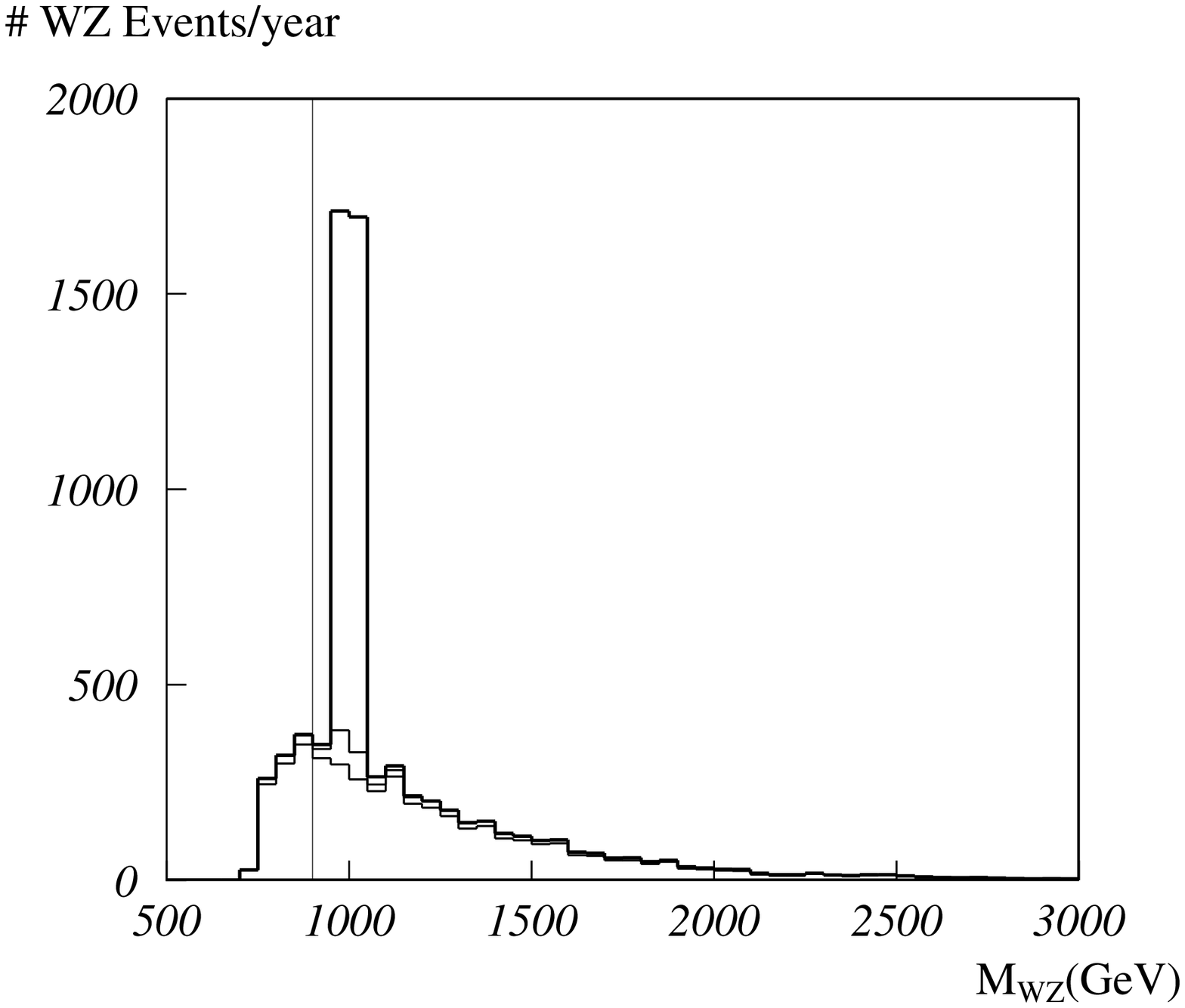}}
\baselineskip=10pt
\smallskip
\noindent
{\bf Fig. 2} - {\it  Invariant mass distribution of the $W^{\pm}Z$ pairs
produced
per year at LHC for $M_V=1000~GeV$, $\gs=22$ and $b=0$. The applied cuts are
$(p_T)_Z>360~GeV$ and $M_{WZ}>900~GeV$. The lower, intermediate and higher
histograms refer the background (3137 events),
background plus fusion signal (3536 events) and
background plus fusion signal plus $q\bar q$ annihilation signal (6294
events).}

Fig. 2 and Fig. 3 give the predictions for invariant $WZ$ mass
and $p_T$ of the $Z$ distributions
for $M_V=1000~GeV$, $\gs=22$ and $b=0$ to which corresponds
a width $\Gamma_V=4~GeV$.

\bigskip
\smallskip
\centerline{
\epsfxsize=8truecm
\epsffile[45 214 551 654]{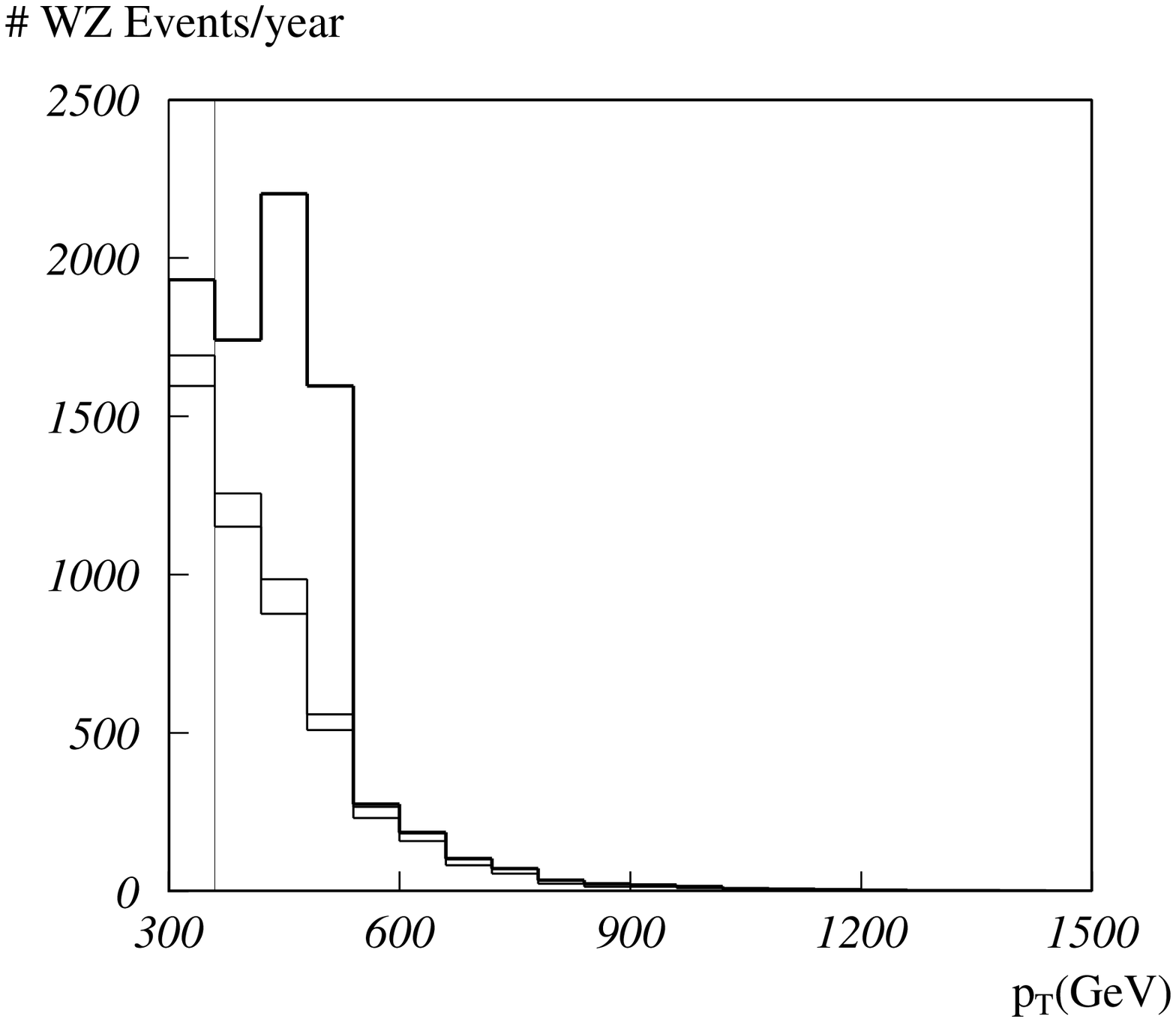}}
\baselineskip=10pt
\smallskip
\noindent
{\bf Fig. 3} - {\it  $(p_T)_Z$ distribution of the $W^{\pm}Z$ pairs
produced
per year at LHC for $M_V=1000~GeV$, $\gs=22$ and $b=0$. The applied cuts
and the number of events  are
the same as in Fig. 2.}
\smallskip

\baselineskip=14pt
Fig. 4 and Fig. 5 give the predictions for invariant $WZ$ mass
and $p_T$ of the $Z$ distributions
for $M_V=1500~GeV$, $\gs=20$ and $b=0.016$ to which corresponds
a width $\Gamma_V=35~GeV$.

The vertical lines in the graphs indicate where the lower cuts in $M_{WZ}$
and $p_T$ have been put for the illustrated cases.
The invariant mass distributions show a peak around
the mass of the $V$, and the $p_T$ distribution
is characterized by a jacobian peak, the broadness
being directly related to the $V$ width.

\smallskip
\centerline{
\epsfxsize=8truecm
\epsffile[45 214 551 654]{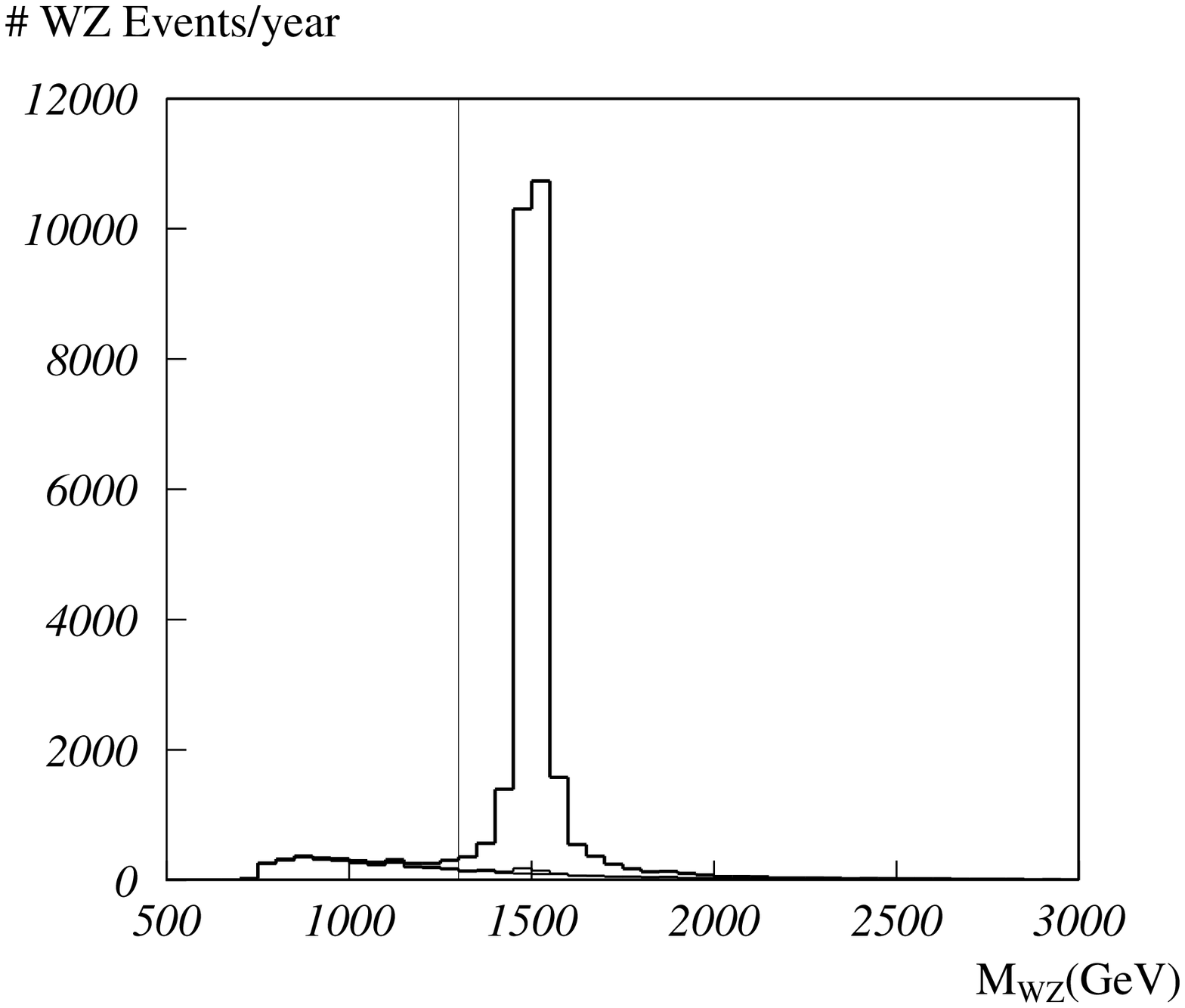}}
\baselineskip=10pt
\smallskip
\noindent
{\bf Fig. 4} - {\it  Invariant mass distribution of the $W^{\pm}Z$ pairs
produced
per year at LHC for $M_V=1500~GeV$, $\gs=20$ and $b=0.016$.
The applied cuts are
$(p_T)_Z>360~GeV$ and $M_{WZ}>1300~GeV$. The lower, intermediate and higher
histograms refer the background (1234 events),
background plus fusion signal (1500 events) and
background plus fusion signal plus $q\bar q$ annihilation signal (27210
 events).}
\smallskip

\newpage
\baselineskip=14pt
For both bosons
decaying leptonically, which is the gold-plated signal,
 one has to multiply by the branching factor
$B(Z\to\ell^+\ell^-)\cdot B(W^\pm\to\ell^\pm{\bar \nu}_\ell)$
$\approx$ 1.5$\%$, for $(\ell=e,\mu)$.
The figures show that, even after multiplying by the branching factor
corresponding to selecting only leptonic decays of $W$ and $Z$, one is left
with a statistically significant
signal having quite well distinguished features both in $M_{WZ}$
and $(p_T)_Z$ distributions.

The sensitivity increases if $WZ$ reconstruction can be done using
the $l\nu,jj$ final state \cite{yuan}.
We can infer that a mass discovery limit for charged vector resonances
around 2 TeV can be achieved at LHC for a large domain of the BESS parameter
space. Nevertheless there are still some parameter values which lead, even
for light $M_V$ masses, to too small number of events to be discovered.

\bigskip
\smallskip
\smallskip
\smallskip
\centerline{
\epsfxsize=8truecm
\epsffile[45 214 551 654]{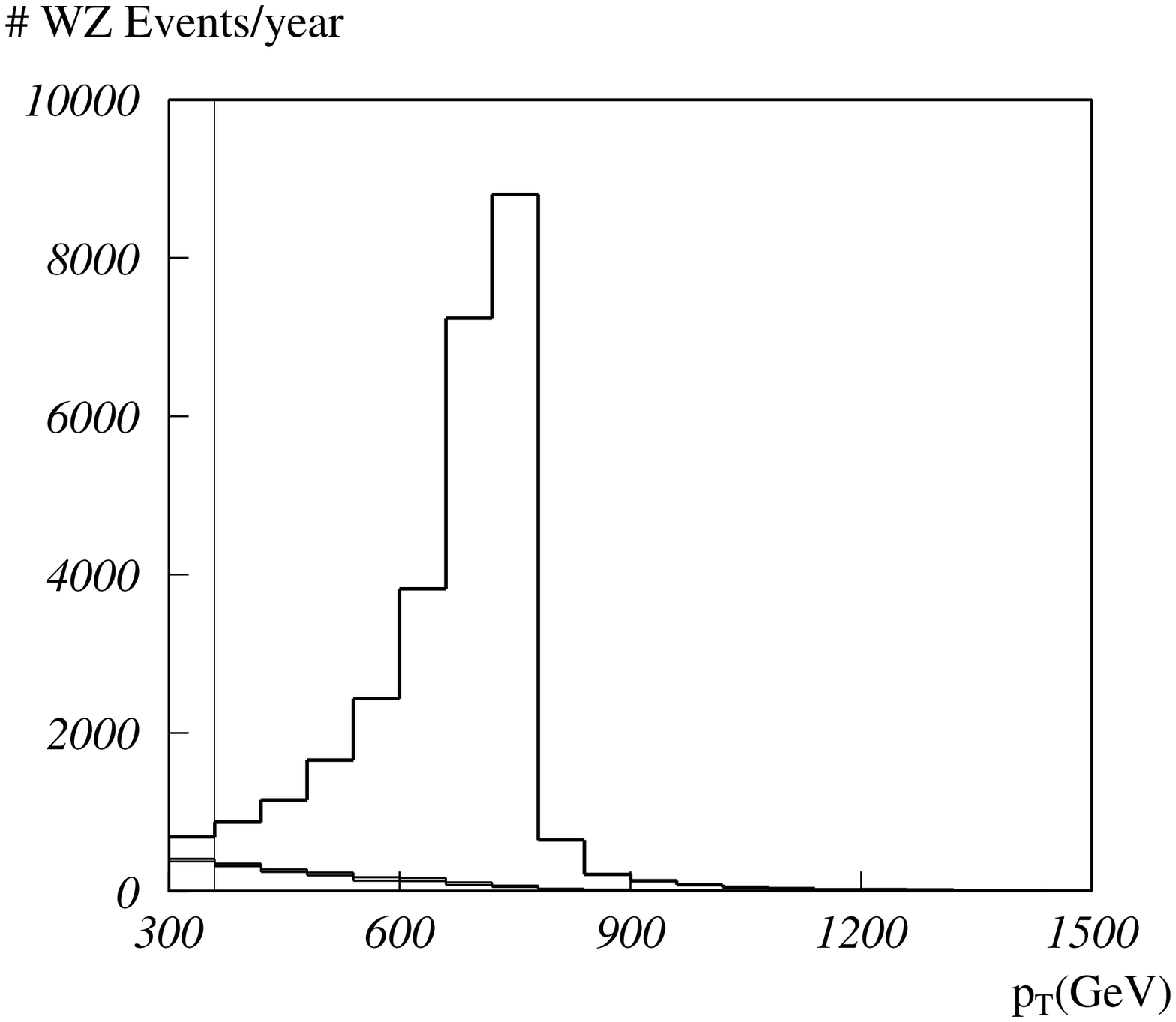}}
\baselineskip=10pt
\smallskip
\smallskip
\noindent
{\bf Fig. 5} - {\it  $(p_T)_Z$ distribution of the $W^{\pm}Z$ pairs
produced
per year at LHC for $M_V=1500~GeV$, $\gs=20$ and $b=0.016$.
The applied cuts and the number of events  are
the same as in Fig. 4.}
\smallskip
\smallskip
\smallskip
\baselineskip=14pt
\resection{Tevatron upgrade}

We have also considered the detection of a signal from strong
electroweak sector at a possible upgrade of the Fermilab
Tevatron. The option, we have chosen, is the one with the
doubling of the c.m. energy of the collider to  4 $TeV$ with
an integrated luminosity
of $10~fb^{-1}$.
In this case we do not consider the second mechanism of
production (the fusion), because its contribution, due to the lower
energy of the collider with respect to LHC,
 is negligible.

We have studied some
examples with different choices of $M_V$, $b$ and $g''$
to give an estimate of the sensitivity of this
option for the upgrading  of the Tevatron.

Fig. 6 and Fig. 7 give the predictions for invariant $W^+Z$ mass
and $p_T$ of the $Z$ distributions
for $M_V=600~GeV$, $\gs=13$ and $b=0.01$ to which corresponds
a width $\Gamma_V=0.9~GeV$.
\newpage
\centerline{
\epsfxsize=8truecm
\epsffile[45 214 551 654]{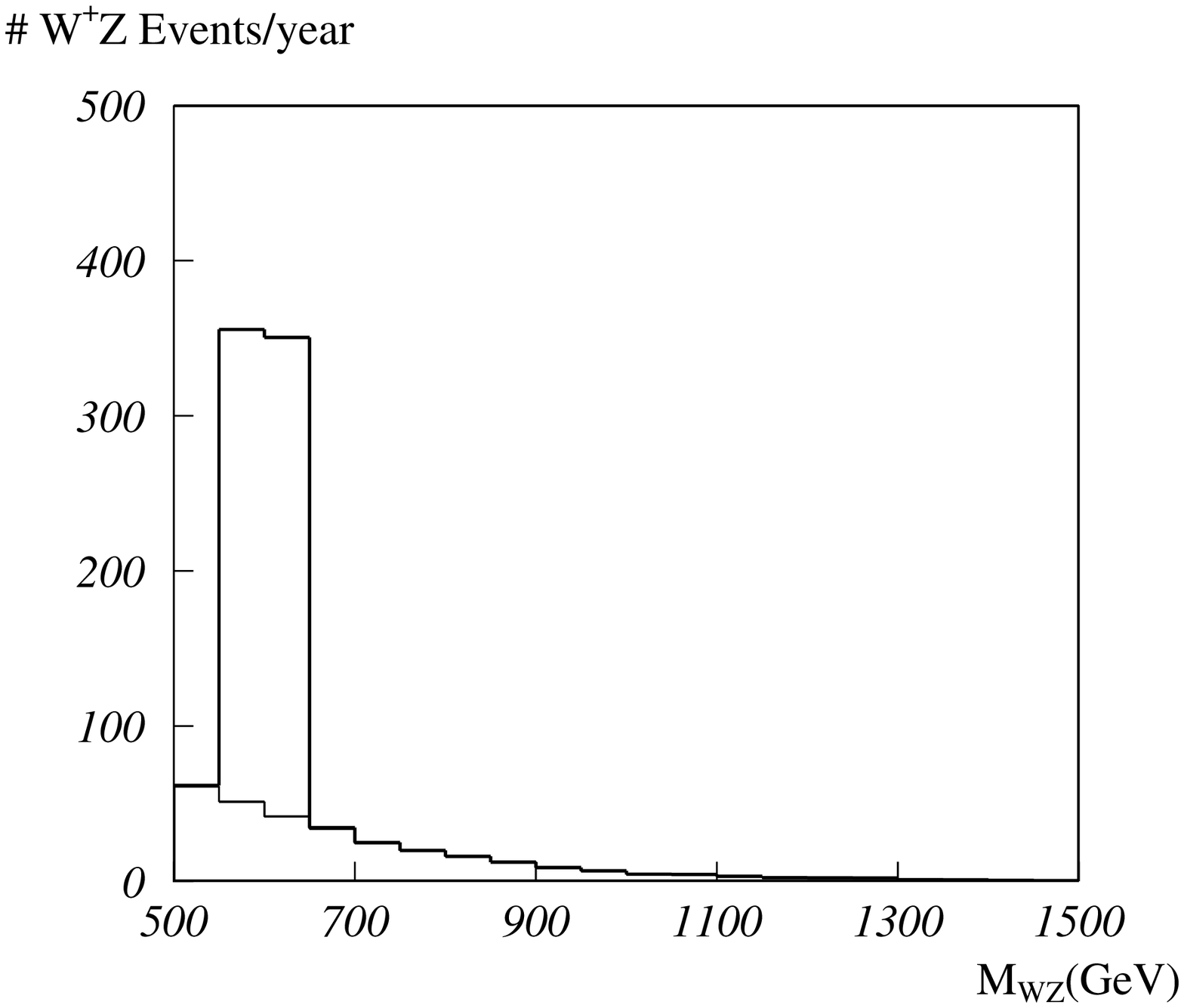}}
\baselineskip=10pt
\smallskip
\smallskip
\smallskip
\noindent
{\bf Fig. 6} - {\it  Invariant mass distribution of the $W^{+}Z$ pairs
produced
per year at Tevatron Upgrade for $M_V=600~GeV$, $\gs=13$ and $b=0.01$.
The applied cuts are
$(p_T)_Z>180~GeV$ and $M_{WZ}>500~GeV$. The lower, higher
histograms refer the background (296 events),
background plus $q\bar q$ annihilation signal
(912 events).}
\smallskip
\smallskip
\bigskip

\baselineskip=14pt
The signal is doubled by adding the $W^-Z$ channel
final state.
Even after multiplying by the appropriate branching ratio one is
left with a statistical significant signal.

\smallskip
\bigskip
\smallskip
\smallskip
\smallskip
\centerline{
\epsfxsize=8truecm
\epsffile[45 214 551 654]{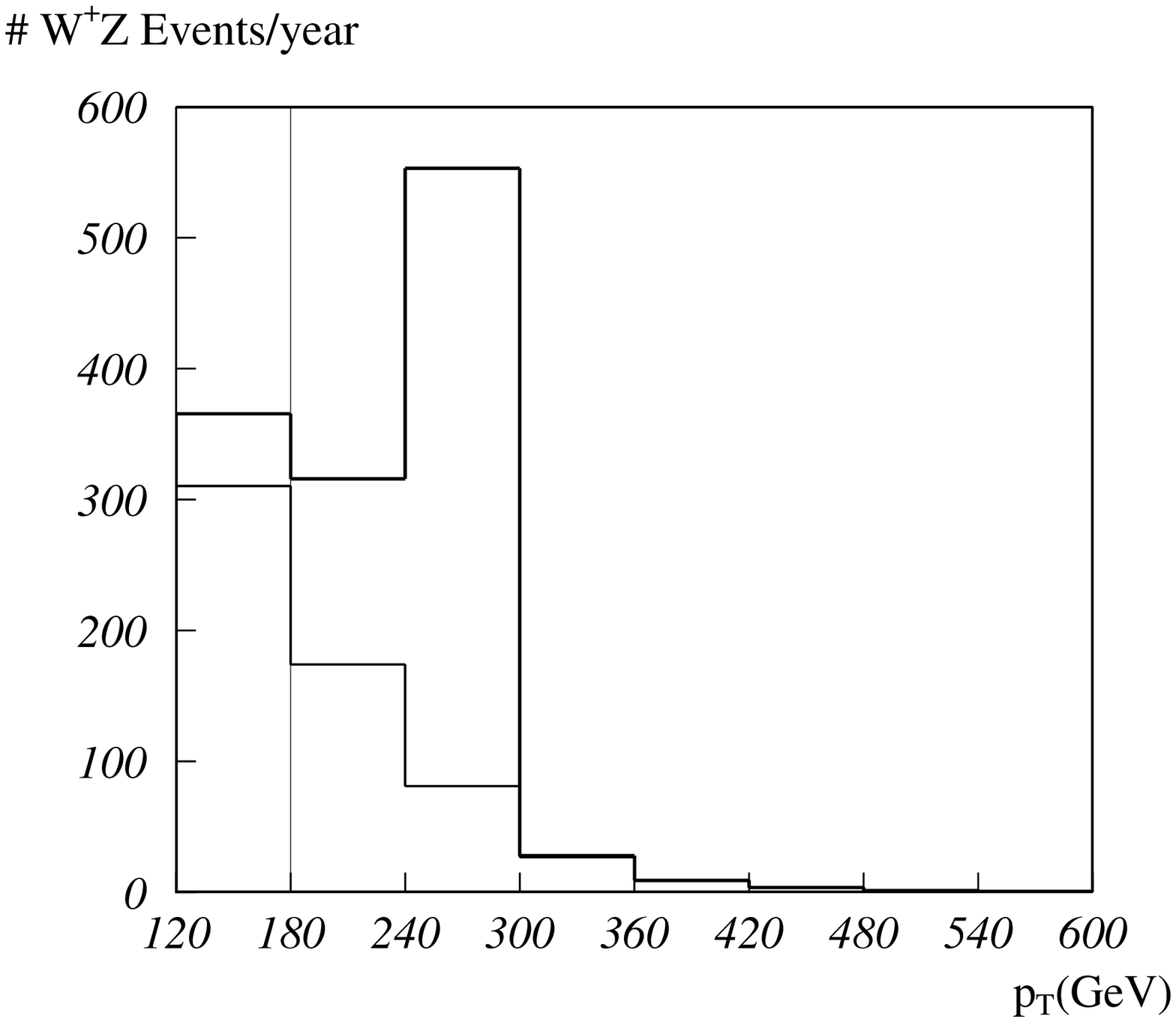}}
\baselineskip=10pt
\smallskip
\noindent
{\bf Fig. 7} - {\it  $(p_T)_Z$ distribution of the $W^{+}Z$ pairs
produced
per year at Tevatron for $M_V=600~GeV$, $\gs=13$ and $b=0.01$.
The applied cuts and the number of events are
the same as in Fig. 6. }
\smallskip
\smallskip
\smallskip
\bigskip

\baselineskip=14pt
As shown by Fig. 8 and Fig. 9, results are depending significantly on the
values
of BESS parameters $b$ and $\gs$. For the same $V$ mass the
case corresponding to the choice $\gs=20$ and $b=0.016$
leads to roughly five times more events. Increasing the mass
to $800~GeV$ reduces the signal by roughly a factor of five.
In a definite region of the parameters $(b,g/\gs)$, the discovery limit of the
Tevatron Upgrade can reach masses $M_V\sim 1~TeV$.

\smallskip
\centerline{
\epsfxsize=8truecm
\epsffile[45 214 551 654]{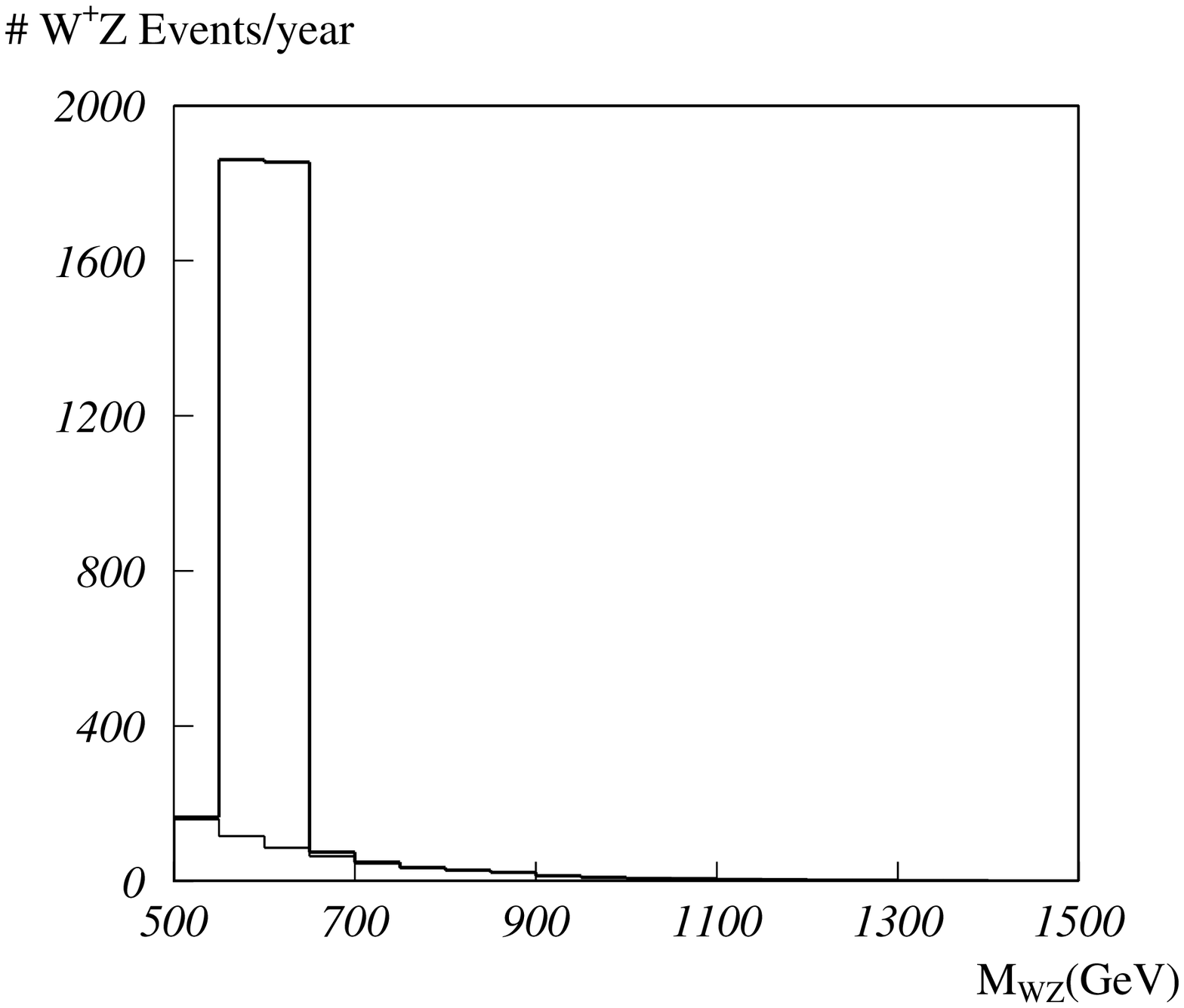}}
\baselineskip=10pt
\smallskip
\noindent
{\bf Fig. 8} - {\it  Invariant mass distribution of the $W^{+}Z$ pairs
produced
per year at Tevatron Upgrade for $M_V=600~GeV$, $\gs=20$ and $b=0.016$.
The applied cuts are
$(p_T)_Z>120~GeV$ and $M_{WZ}>500~GeV$. The lower, higher
histograms refer the background (606 events),
background plus $q\bar q$ annihilation signal
(4142 events).}
\smallskip

\baselineskip=14pt

\smallskip
\centerline{
\epsfxsize=8truecm
\epsffile[45 214 551 654]{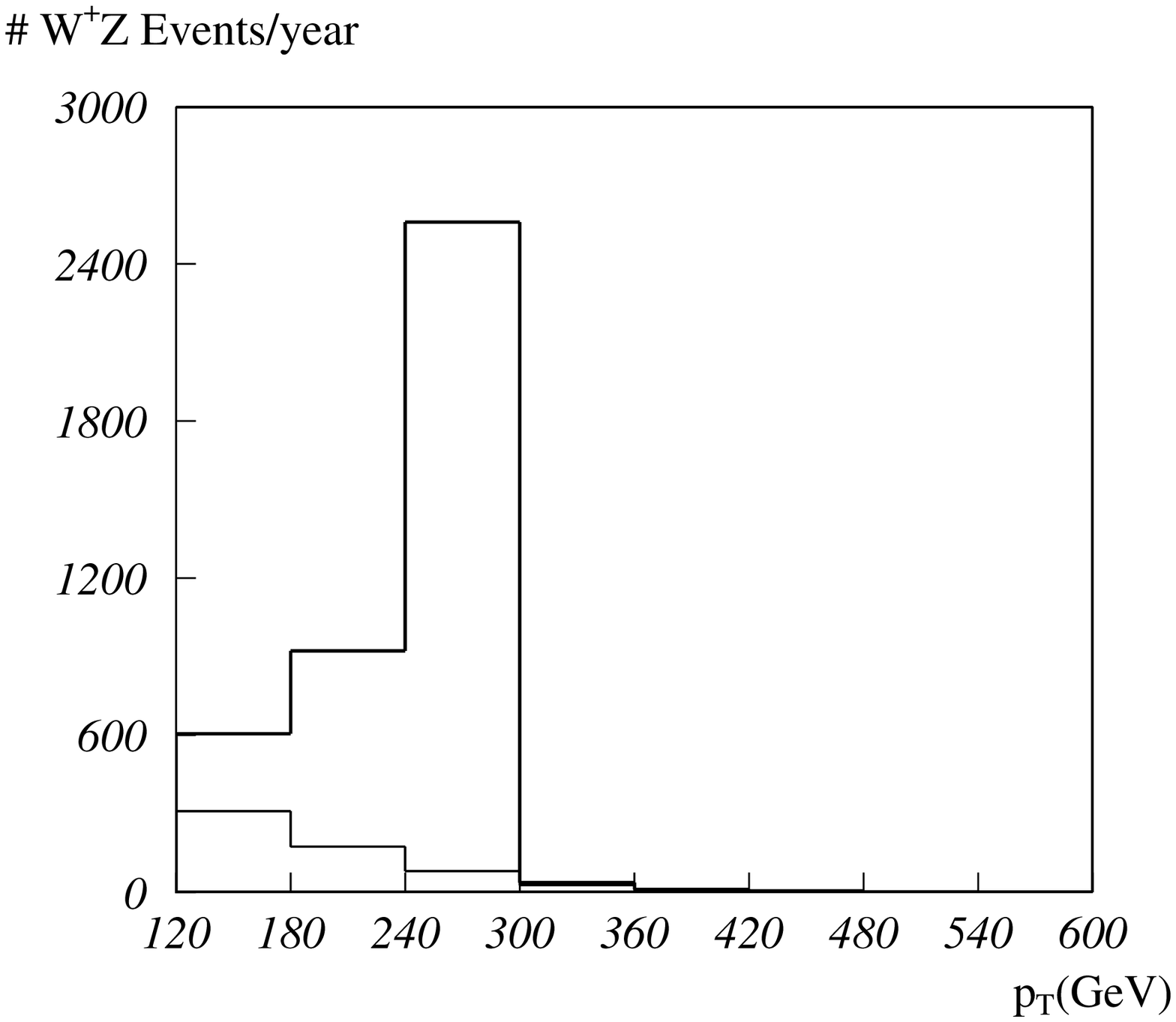}}
\baselineskip=10pt
\smallskip
\noindent
{\bf Fig. 9} - {\it  $(p_T)_Z$ distribution of the $W^{+}Z$ pairs
produced
per year at Tevatron for $M_V=600~GeV$, $\gs=20$ and $b=0.016$.
The applied cuts and the number of events are
the same as in Fig. 8. }
\smallskip

\baselineskip=14pt

\resection{$e^+e^-$ colliders}

Future $e^+e^-$ colliders are sensitive to the neutral
$V^0$ resonance  if the mass
 $M_V$ of the new boson multiplet lies not
far from the maximum machine energy, or if it is lower,
such a resonant contribution would be quite manifest.
The result of our analysis \cite{lin}
is that also virtual effects are important.
It appears that annihilation into a fermion pair in such  machines, at the
considered luminosities, would marginally improve on existing limits
if polarized beams are available and left-right asymmetries are measured.
On the other hand, the process of $W$-pair production by $e^+e^-$
annihilation would allow for sensitive tests of the
strong sector,
especially if the $W$ polarizations are  reconstructed from their
decay distributions, and the more so the higher the energy of the machine.
This is because BESS modifies the standard couplings  in such
a way that the typical cancellations present in the SM do not happen
anymore and the amplitude is growing with $s$.

If the masses  of the $V$ bosons are  higher than the maximum c.m.
energy,
they give rise to
indirect effects in the
$e^+e^-\rightarrow f^+f^-$ and  $e^+e^-\rightarrow W^+W^-$  cross sections.

We have analyzed
cross-sections and asymmetries for the channel $e^+e^-\rightarrow f^+f^-$
and $e^+e^-\rightarrow W^+W^-$.
For the purposes of our calculation we have also assumed that it will
be possible to separate
$e^+e^-\rightarrow W^+_L W^-_L$, $e^+e^-\rightarrow W^+_L W^-_T$, and
$e^+e^-\rightarrow W^+_T W^-_T$. The distribution of the $W$ decay angle
in its c.m. frame depends indeed in a very distinct way from its helicity,
being peaked forward (backward) with respect to the production direction
for positive (negative) helicity or at $90^o$ for zero helicity.

We consider the $WW$ channel, for one $W$  decaying leptonically
and the other hadronically.
To discuss the restrictions on the parameter space for masses of the
resonance a little higher than the available energy we have taken into
account the experimental efficiency. We have assumed
an overall detection efficiency of 10\% including
account the branching ratio
$B=0.29$ and the loss of luminosity from
 beamstrahlung.

For a collider at
$\rs=500~GeV$ with an integrated luminosity
of $20~fb^{-1}$ the results are illustrated in Fig. 10. The contours have
been obtained by taking 18 bins in the angular region restricted by
$|\cos\theta|< 0.95$. This figure illustrates the 90\% C.L. allowed regions
for $M_V=600~GeV$
obtained by considering the unpolarized $WW$ differential cross-section
(dotted line), the $W_LW_L$ cross section (dashed line),
and the combination of the left-right asymmetry with all the
differential cross-sections for the different final $W$ polarizations
(solid line). We see that
already at the level of the
unpolarized cross-section we get important restrictions
with respect to LEP1.

\smallskip
\centerline{
\epsfxsize=8truecm
\epsffile[69 263 498 700]{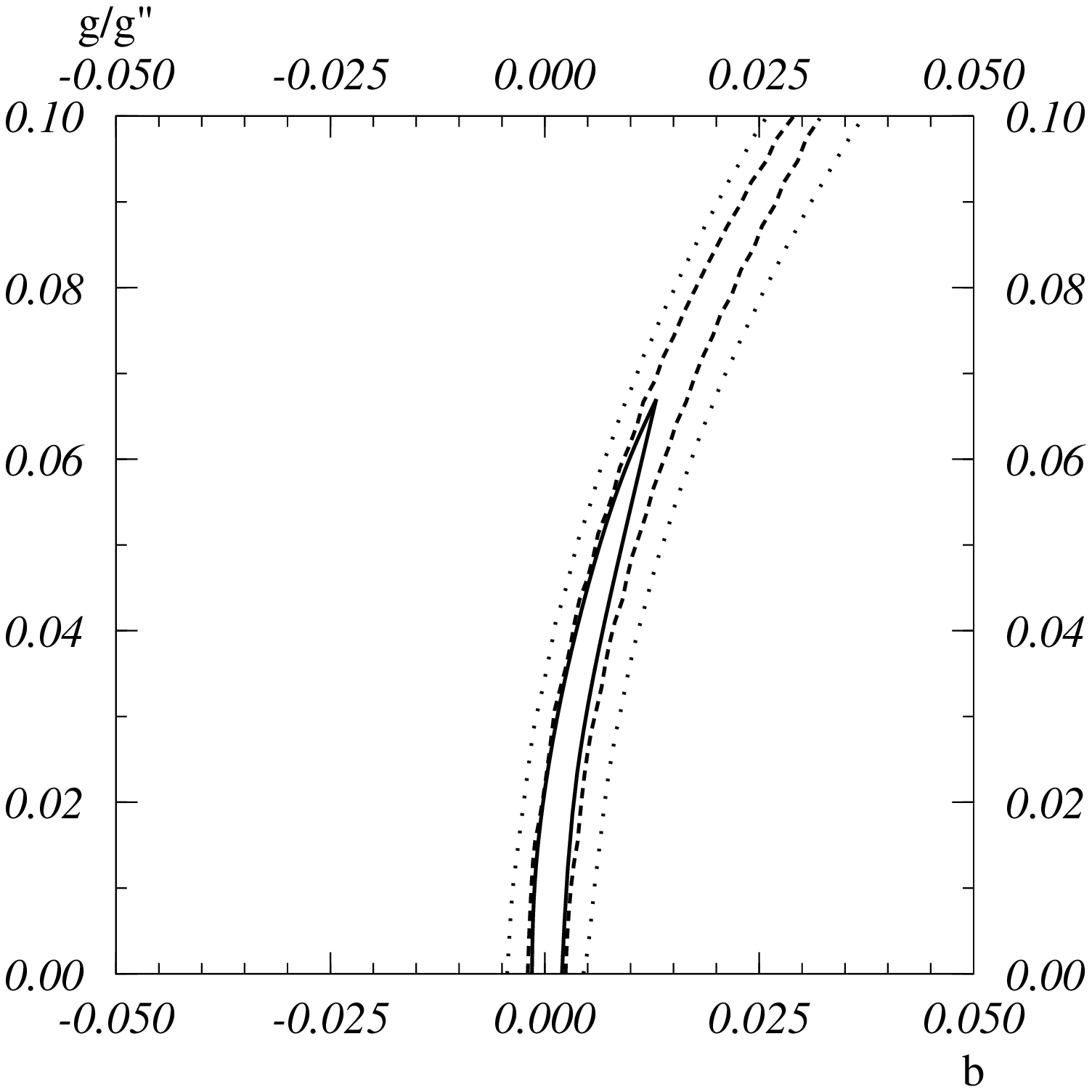}}
\baselineskip=10pt
\smallskip
\noindent
{\bf Fig. 10} - {\it  90\% C.L. allowed regions
for $M_V=600~GeV$
obtained by considering the unpolarized $WW$ differential cross-section
(dotted line), the $W_LW_L$ cross section (dashed line),
and the combination of the left-right asymmetry with all the
differential cross-sections for the different final $W$ polarizations
(solid line).}
\smallskip

\baselineskip=14pt
For colliders with $\rs=1,~2~TeV$
and for $M_V=1.2~{\rm and}~2.5~TeV$ respectively,
the allowed region, combining all the observables,
 reduces in practice to a line.
Therefore, even the unpolarized $WW$ differential
cross section measurements
can improve the bounds.

\resection{Conclusions}

We have used the BESS model, as a rather general frame based on custodial
symmetry and gauge invariance, to examine the possibilities offered by
the upgraded Tevatron, by LHC, and by $e^+e^-$ colliders in the $TeV$ range,
 to test for strong
electroweak breaking.
We have first presented the existing limits on the BESS parameters
 from the latest LEP data, atomic parity violation, and from the ratio
of $W$ to $Z$ mass.
We have then summarized the projections for observables in $WZ$ production
at LHC within present BESS limitations, and compared them with the standard
model backgrounds, within suitable cuts and detection limits.
We have then examined the corresponding observables at the upgraded
Tevatron.
Finally we have discussed the sensitivity to BESS parameters of $TeV$ $e^+e^-$
colliders for annihilation into fermions or into $W$ pairs, considering
cross-sections and asymmetries.
Our study shows the interest of these future facilities in relation to
a possible strong electroweak breaking.


\begin{thebibliography}{99}


\bibitem{BESS}
R. Casalbuoni, S. De Curtis, D. Dominici and R. Gatto,
       Phys. Lett. {\bf B155}  (1985) 95; and
       Nucl. Phys. {\bf B282} (1987) 235



\bibitem{bando} M. Bando, T. Kugo, S. Uehara, K. Yamawaki and T.
Yanagida,  Phys. Rev. Lett.  {\bf 54} (1985) 1215;
A.P. Balachandran, A. Stern and G. Trahern,  Phys. Rev. {\bf
D19 } (1979) 2416.


\bibitem{farhi} For a review see E. Fahri and L. Susskind,  Phys.
Rep. {\bf 74} (1981) 277.

\bibitem{self}   R. Casalbuoni, S. De Curtis, D. Dominici,
F. Feruglio and R. Gatto,
   Phys. Lett. {\bf B258}  (1991) 161.

\bibitem{anich} L. Anichini, R. Casalbuoni, S. De Curtis,
 Univ. di Firenze Preprint,
 DFF-210/10/1994.

\bibitem{alta} G. Altarelli and R. Barbieri, Phys. Lett. {\bf
B253} (1991) 161;
G. Altarelli, R. Barbieri and S. Jadach,  Nucl. Phys. {\bf
B369} (1992) 3;
D.C. Kennedy and P. Langacker,  Phys. Rev. {\bf D44} (1991)
1591.

\bibitem{alta2} G. Altarelli, CERN preprint, CERN-TH 7464/94 October 1994.

\bibitem{josa} I. Josa, T. Rodrigo and F. Pauss,
   CERN 90-10, volume II, p. 796, Proceedings
 of Large Hadron Collider Workshop, Aachen, 4-9 October 1990, Eds. G. Jarlskog
 and D. Rein.

\bibitem{LHC}
R. Casalbuoni, P. Chiappetta, S. De Curtis,  F. Feruglio, R. Gatto,
B. Mele and J. Terron, Phys. Lett. {\bf B249} (1991) 130.

\bibitem{pseudo}
R. Casalbuoni, P. Chiappetta, A. Deandrea, S. De Curtis,  D. Dominici
and R. Gatto, Zeit. f\"ur Phys., {\bf C65} (1995) 327.

\bibitem{yuan} C.P. Yuan, in Perspectives in Higgs Physics, Ed. G. Kane,
World Scientific, 1993.

\bibitem{lin}
R. Casalbuoni, P. Chiappetta, A. Deandrea, S. De Curtis,  D. Dominici
Zeit. f\"ur Phys., {\bf C60} (1993) 315.

\end{thebibliography}
\end{document}